# 2D MoSe$_2$ as a Promising Chemo-resistive Sensor for N$_2$O Detection: A DFT Approach

**Neha Mishra and Bramha P Pandey**

*Department of Electronics and Communication Engineering, MMMUT, Gorakhpur (U.P.), India*.

We have investigated the impact of toxic N$_2$O gas upon structural and electronic properties of 2D MoSe$_2$ monolayer using DFT approach. In this work, as a result of N$_2$O gas absorption, charge transfer, band gap and density of states (DOS) are changed and these parameters are extracted as electronic properties of 2D MoSe$_2$ monolayer nano-gas sensor. Moreover, the band gap is tunable upon the N$_2$O gas absorption for pristine and adsorbed MoSe$_2$ in both Top Mo (Top Se) structures. Also, the spin up and down states in the DOS results considerable magnetic moment altering the magnetic property of the 2D MoSe$_2$ monolayer. Later, the desorption property of the 2D MoSe$_2$ monolayer towards the target N$_2$O gas molecule at three different temperature are calculated. Thus, the paper concludes with outcomes of the structural and electronic properties aligning its behavior as a chemo-resistive nano-gas sensor and showing it as a potential applicant for sensing of toxic N$_2$O gas molecule. The nature of toxic N$_2$O gas molecule is not explored till date using 2D monolayer, thus, in this work it is estimated through simulated results and the experimental verification is awaited.

*Keywords*: DFT, Adsorption, MoSe$_2$, Charge transfer, Recovery time, N$_2$O

1. **Introduction**

The day-to-day increasing environmental disturbances, caused due to the harmful gases are dangerous to human health and society. An effective step towards minimizing these issues could be designing of nano-gas sensors that can effectively detect these gases and predict their behavior. In the recent years, Two-Dimensional (2D) transition metal dichalcogenide (TMD) are widely utilized to implement gas sensors and based devices, due to their unique structural and electronic properties [1, 2, 3]. Some of the popularly used materials reported are MoS$_2$, MoSe$_2$, WS$_2$ and WSe$_2$ due to their tunable band gap which modulates their electronic and optoelectronic properties giving an advantage over other classes of materials [4]. They share similar structure as graphite but non zero intrinsic band gap that is the root cause for varying properties. The Van der Waals (vdW) interaction is responsible to bind the gas molecule to the surface of 2D monolayer which is frequently known to be physical adsorption while in chemical adsorption chemical bonding or ionic bonding dominates over vdW interaction [5,6,7]. To obtain better performance of nano-gas sensors based on 2D materials, it is required to modulate their properties by developing new design strategies [8, 9, 10]. For effective gas sensing, large surface to volume ratio and surface decoration is adopted now a days to alter the properties of the 2D MoSe$_2$ monolayer [11, 12]. The structural, electronic and magnetic properties of the 2D monolayer are changed upon the absorption of gas molecule [13, 14]. The electronic property comprises of band gap, density of states and recovery time. The band gap of 2D materials can easily be tuned between 1.0-2.5 eV by changing the number of inter layers due to quantum confinement effect [15,16] as well as by numerous other techniques reported by other workers [17,18]. On the other hand, density of state (DOS) is responsible for the redistribution of electrons near the conduction band as a result of absorption of gases on the surface of 2D material. It also conveys about the Fermi level changes post and prior absorption of gases upon the 2D material [19, 20]. The symmetry/asymmetry in DOS behavior reveals the magnetic property of the 2D material. The total magnetic moment ($\mu_B$) is contributed due to the spin up and down states of the atoms of the 2D material [21]. Further, sensitivity is an important parameter for gas sensing. It is reported that $MoS_2$ gas sensors show high sensitivity towards NO and $NO_2$ gas to a lower detection limit to 0.8 ppm and 20 ppb respectively [22, 23]. It is highly dependent on the adsorption strength of the target gas molecule over the surface of the 2D material [24]. Numerous potential applications of these materials have been explored, such as catalysis [25], photo emitting devices [26], hydrogen storage and Li-ion batteries [27]. To further increase the application of 2D material, hetero structures are analyzed. They are formed of vdW interactions unless of covalent bonding. Also, small desorption time of these gases allows them to come under the category of disposable gas sensors [28,29].Thus, all these prodigious properties of 2D TMDs help in designing of nano-gas sensors with high sensitivity and stability.

Nitrogen oxides like nitrogen dioxide ($NO_2$), nitrogen monoxide (NO) and nitrous oxide ($N_2O$) are undesirable components for both human health and environment. Firstly, for the human health, nitrous oxide ($N_2O$) is *N*-methyl-D-aspartate (NMDA)-antagonist drug which is a cause for neurotoxicity of brain. It is also associated to cardiovascular problems arising due to exposure to this gas. Long term effect of this gas includes increased homocysteine levels leading to neuronal death. Several Investigators have found that different brain conditions are varying reactions to each form of toxicity, with neonatal brain more prone to NMDA antagonism and damaged brain more prone to changes in cerebral blood flow [30, 31]. Secondly, from the aspect of environmental pollution, it is a dominant greenhouse gas and cause of ozone layer depletion. It is also found to be fuel additive in nature. A major portion of these emissions originate from the bacterial and fungal respiratory processes in soil, biologically known as de-nitrification and nitrification respectively [32]. The detection of the detrimental impact of $N_2O$ gas using 2D $MoSe_2$ monolayer is a point of interest in this work.

In this work, the adsorption of $N_2O$ gas on surface of 4x4x1 $MoSe_2$ monolayer has been studied and investigated the $N_2O$ gas absorption impact on electronic properties such as density of states (DOS), band gap and recovery time along with the magnetic moment of the $MoSe_2$ monolayer. The paper is distributed into four sections, section 1 introduces about the 2D material and its implementation in gas sensor whereas section 2 briefs about the computational details used in simulation thereafter, in section 3 the extracted parameters from calculations are discussed and thus section 4 concludes the work with the findings and scope of improvement in future.

**2. Computational Details**

In this work, DFT calculations have been performed using SIESTA [33]. Perdew–Burke–Ernzerhof (PBE) [34] with Generalized Gradient Approximation (GGA) [35] method is opted to account into the correlation and exchange of electrons along with conjugate gradient [36] as a method of relaxation to obtain the most optimized structure of the studied system. The $4X4X1$ super cell was allowed to relax until the total ground state energy was less than $10^{-4}$ eV/atom and maximum force is reduced below $0.055$ eV/Å. The cut-off energy was set to $350$ eV. For effective sampling, a $5X5X1$ Monkhorst-Pack grid [37] in the k-point sampling of the Brilliouin zone is used. To maintain periodicity in two directions, a vacuum of 12 Å is applied to z-axis to avoid interactions between two adjacent slabs. To include the van der Waal interaction between the target gas molecule and monolayer, DFT-D2 method by Grimme [38] is adopted. The Hirshfeld charge analysis is performed as a method to describe the effective charge transfer [39, 40].

## 3. Results and Discussion

*3.1. Adsorption of $N_2O$ gas and impact on magnetic moment of 2D $MoSe_2$ monolayer*

After the complete optimization of the structure, two possible configurations are presented: Top Mo and Top Se monolayer respectively where $N_2O$ gas is adsorbed at varying distances as shown in the Fig.1 (a) (b). In order to study the behavior of 2D $MoSe_2$ layer towards the absorption of $N_2O$ gas molecule is investigated. The adsorption energy ($E_{ads}$) parameter is studied at different positions corresponding to configurations (Top Mo, Top Se) are reported in this work. Likewise, for the Top Mo configuration, during the adsorption energy analysis, the bond length between Mo-Se and N-Mo atom and the bond angle between Se-Mo-Se are found to be shortened from the pristine phase as a consequence of the absorption of gas molecule depicted in the Table 1. Similarly, for the Top Se configuration, the bond length between Mo-Se and N-Se with bond angle between Se-Mo-Se in pristine phase and adsorbed phase is investigated separately and enlisted in Table 1. It is noticed that decrease in bond length is more pronounced for N-Se than N-Mo in Top Se and Top Mo respectively. This witnesses the reason for increase in charge transfer and adsorption energy in Top Se configuration than its Top Mo counterpart. The change of the bond length of $N_2O$ gas molecule is a symbolic of its activation towards the 2D $MoSe_2$ monolayer [41]. The adsorption energy ($E_{ads}$) signifies the extent to which the gas molecule adsorbs on the 2D $MoSe_2$ monolayer which is governed by the following equation [41]:

$$E_{ads} = E_{gas+monolayer} - E_{gas} - E_{monolayer} \quad (1)$$

A value of $E_{ads} < 0$ indicates that $N_2O$ gas molecule gains electrons from the 2D $MoSe_2$ monolayer with activated N atom acting as localization center whereas $E_{ads} > 0$ implies 2D $MoSe_2$ monolayer is accepting electrons from the $N_2O$ gas molecule with monolayer acting as localization center. The order of the adsorption energies for Top Mo and Top Se configurations are -1.05 < 1.08 <1.55 eV and 1.38 < 3.00 < 7.67 eV, respectively as listed in the Table 1 and the decrement behavior of the adsorption energy with different absorbent distance towards the 2D $MoSe_2$ monolayer is shown in Fig.1 (c). Interestingly, the adsorption energy increases with the decrease in the distance between the $N_2O$ gas molecule and 2D $MoSe_2$ monolayer showing higher interaction and stability that is in alignment with the reported work [41]. Thus, it is clearly noticed that the order of adsorption energies (/atom) supports strong chemisorption as the adsorption distance is less than 3 Å and energy range is greater 50-500 meV/atom which is justification with the reported work [42]. Hence, it is observed that Top Se configuration is having higher adsorption energy and magnetic moment than its counterpart which is desirable for an efficient nano-gas sensor.

**Table 1**
Values of Adsorption energy, Fermi Energy, Bond Distance and Magnetic Moment with varying $N_2O$ gas distance on 2D $MoSe_2$ monolayer for Top Mo and Top Se configuration.

| System | Adsorbent Distance (D) | Adsorption energy $E_{ads}$ (eV) | Bond Distance (Å) | | | | Fermi Energy $E_F$ (eV) | Magnetic Moment $(\mu_B)$ |
|---|---|---|---|---|---|---|---|---|
| | | | N-Mo | | O-Se | | | |
| | | | Pristine | Adsorbed | Pristine | Adsorbed | | |
| **Top Mo** | 2 | -1.05 | 3.59 | 1.90 | 5.95 | 3.15 | -4.13 | 1.69 |
| | 1.5 | 1.08 | 3.09 | 1.63 | 5.45 | 2.88 | -4.21 | 1.54 |
| | 1 | 1.55 | 2.59 | 1.37 | 4.95 | 2.62 | -4.22 | 1.45 |
| | | | N-Se | | O-Se | | Fermi Energy $E_F$ (eV) | Magnetic Moment $(\mu_B)$ |
| | | | Pristine | Adsorbed | Pristine | Adsorbed | | |
| **Top Se** | 2 | 1.38 | 2.00 | 1.06 | 4.36 | 2.31 | -4.33 | 1.76 |
| | 1.5 | 3.00 | 1.51 | 0.80 | 3.86 | 2.04 | -4.17 | 1.48 |
| | 1 | 7.67 | 1.01 | 0.53 | 3.36 | 1.78 | -3.99 | 0.63 |

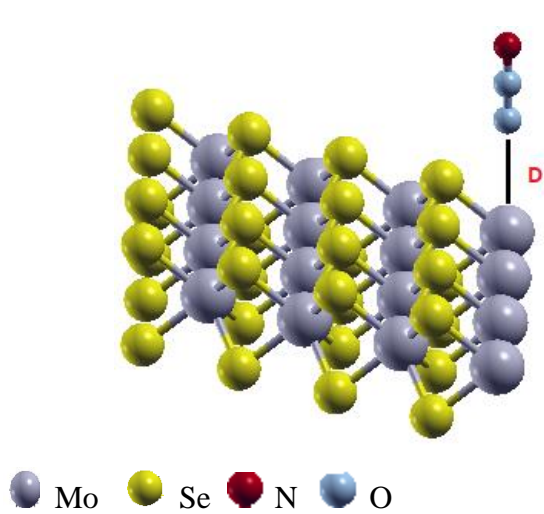 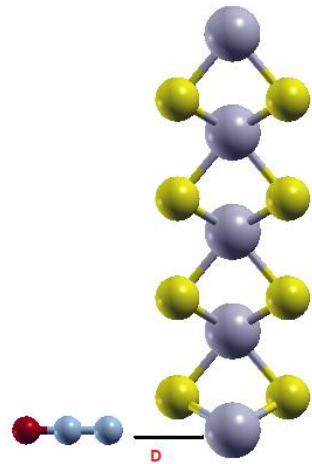

(a)

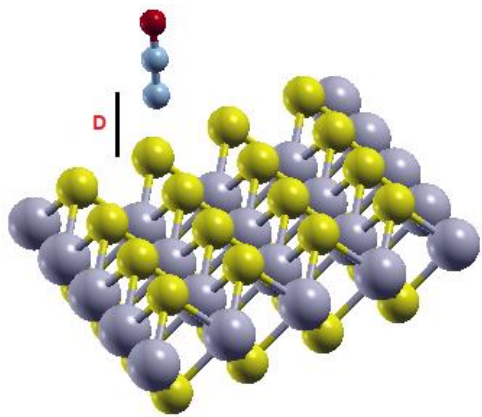 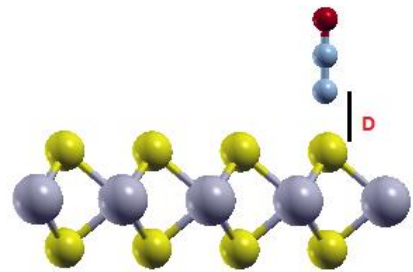

(b)

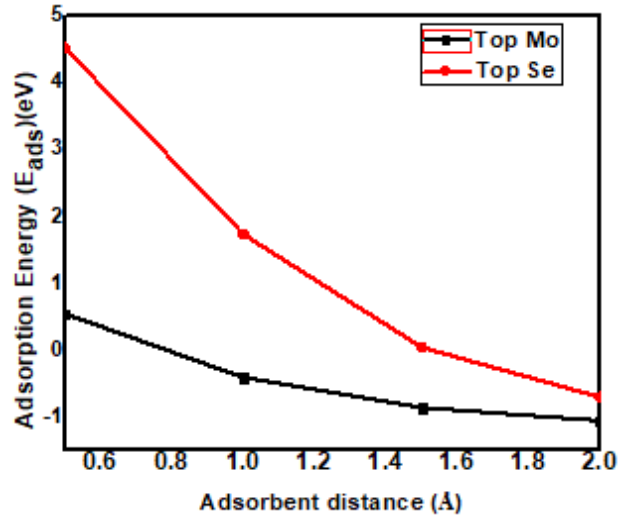

(c)

Fig. 1 Schematic of optimized N₂O gas molecule on (a) Front view and side view of Top Mo configuration at distance D (2 to 1Å) (b) Front view and side view of Top Se configuration at distance D (2 to 1 Å ) (c) Adsorption energy variation as a function of adsorbent distance.

*3.3. Band structure and DOS Analysis of N₂O adsorption on 2D MoSe₂ monolayer*

The impact of N₂O adsorption on the electronic properties of 2D MoSe₂ monolayer can be analyzed from the DOS diagram as portrayed in the Fig. 2 (a)-(d). The density of states for Top Mo and Top Se configuration respectively is studied in this section. Apart from the comparison of adsorption energy ($E_{ads}$) and charge transfer ($Q_T$) to predict the chemisorption or physisorption, the assessment of distance between the activated N atom and transition metal (TM) Mo (Se) atom for Top Mo (Top Se) configurations with the sum of their covalent radii is another criteria to identify the nature of adsorption [43]. The sum of covalent radii being larger than the distance between the activated atom and TM result into chemisorption and vice versa is applied for physisorption [44, 45]. The covalent radii sum for N atom and TM Mo (Se) atom is 2.24 Å (1.87 Å) that is larger than the distance between N-Mo (N-Se) atom (Table 1) confirming the chemisorption for both Top Mo (Top Se) configurations respectively.

For the Top Mo configuration, it can be inferred from the Fig. 2 (a) that there is a wider gap near the Fermi level of pure MoSe₂ that proves its semiconducting nature. Also, with the adsorption of N₂O gas on the MoSe₂ monolayer the gap is narrowed due to upward shift of the Fermi level as shown in the Fig. 2 (a). As a result the Fermi level rises up by -1.76 eV clearly illustrating the contribution of the N₂O gas on the MoSe₂ monolayer. We can see from Fig.2 (b)-(d) that peaks in the DOS of the N₂O system is shifted both in spin up and spin down showing the asymmetry from 2 to 1 Å and contributing to a magnetic moment as shown in Table 1 which is due the adsorption of the N₂O gas molecule. This arises due to the redistribution of the electrons from the adsorption of the N₂O gas. The DOS of the adsorbed system shifts towards the left as compared to the pure MoSe₂ system which is caused by the loss of electron from the N₂O molecule to the MoSe₂ monolayer. Thereafter, based on the partial DOS analysis of Top Mo configuration from Fig. 2 (b), every peak of the 2p orbital of N atom is completely overlapped with the 4d orbital of the Mo atom, implying that there is strong hybridization with the N atom thus giving rise to strong binding force for the Mo-N bond.

Interestingly, for the Top Se configuration, from the Fig. 2 (c) the peaks of the DOS of the $N_2O$ system are overlapping less with pristine $MoSe_2$ as related to Top Mo configuration, showing less activation of the gas molecule towards monolayer. This asymmetrical nature of spin states (up and down) give rise to a magnetic moment as shown in Table 1. The DOS of the adsorbed system shifts towards the right as compared to the Top Mo system. There is a decline in the peak of the DOS for the $N_2O/MoSe_2$ system compared to the pure $MoSe_2$ resulting from the adsorption of the gas molecule. On the other hand, looking in to PDOS in Fig. 2 (c) there exists large overlapping between N-2p orbital and Se- 4p orbital compared to Mo-4d orbital showing strong hybridization which is the reason for the improved electronic properties. This results in formation of additional N-Se bonds during adsorption when the adsorbent distance is decreased from 2 to 1 Å [46]. Again, the wide gap near the Fermi level in the DOS of the pristine $MoSe_2$ confirms its semiconducting properties [46].

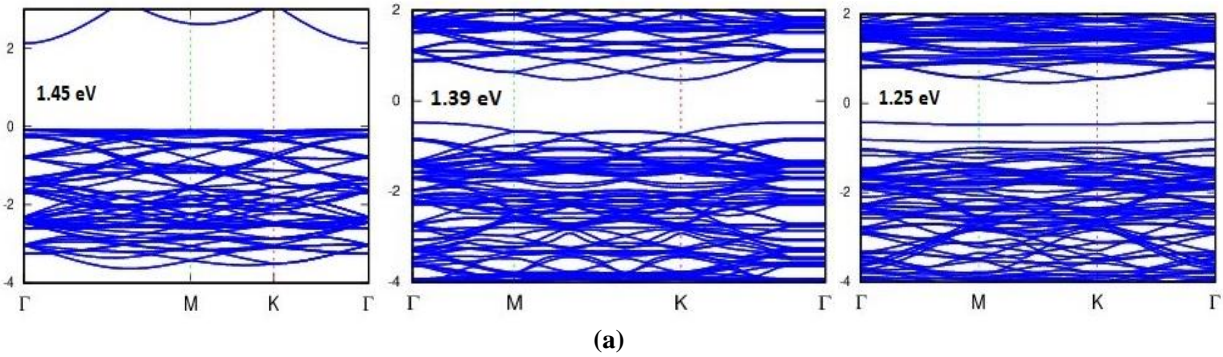
(a)

The band gap is calculated to be 1.45eV for pristine $MoSe_2$ from the TDOS which is in close alignment with the reported value [47] as shown in Fig. 2(a). Similarly, the impact of $N_2O$ gas adsorption on the band gap is clearly observed with 1.39 eV and 1.25eV for Top Mo (Top Se) configurations respectively. Likewise, the same modulation is observed in the total and partial DOS plots for both the configurations as shown in the Fig.2 (b)-(d). Also, strong overlapping between O 2p orbital and Se 4d orbital is the reason for additional O-Se bond during adsorption. Thus, the modulation in band gap is more pronounced for Top Mo configuration due to which its electronic properties are more announced making it favorable candidate for nano-gas sensor.

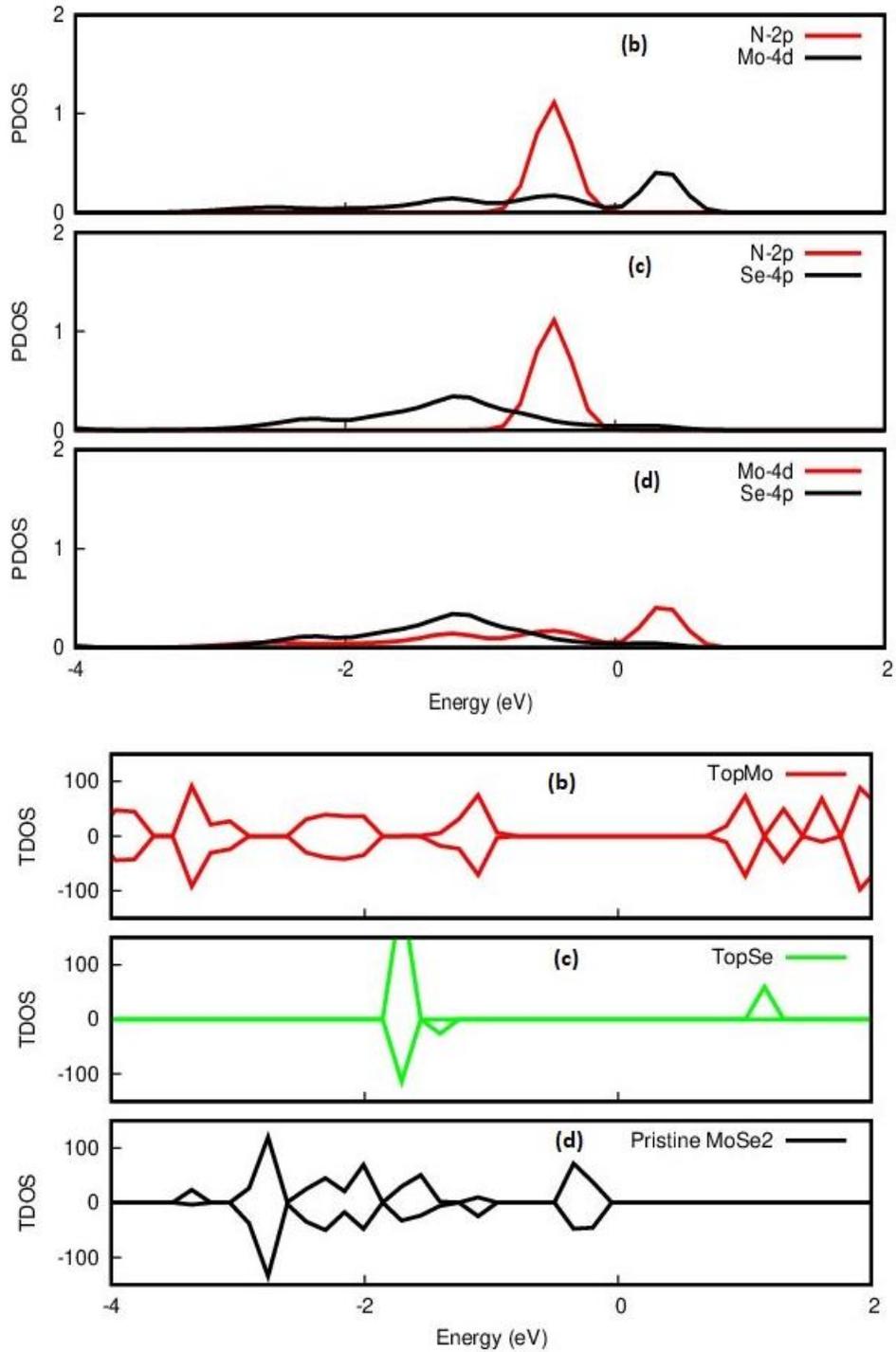

Fig.2 (a) Band Structure of Pristine, Top Mo and Top Se MoSe$_2$ adsorbed system (b) TDOS distribution of the MoSe$_2$ system in Top Mo configuration and Partial DOS of Mo-4d and N-2p orbital (c) TDOS distribution of MoSe$_2$/ N$_2$O system in Top Se configuration and Partial DOS of N-2p and Se-4p orbital. d) TDOS distribution of MoSe$_2$/ N$_2$O system in pristine configuration and Partial DOS Mo-4d of Se-4p orbital. Fermi level shifted to zero.

*3.4 Charge Transfer and Selectivity of N$_2$O adsorbed 2D MoSe$_2$ monolayer*

There are various sensing mechanism implemented for different sensor systems: charge transfer between a transducer and an analyte, change in the Schottky barrier altitude when the gas is adsorbed upon and so on [46]. To explicate further, sensing mechanism using charge transfer ($Q_T$) is focused in this section. Also, estimation of induced electrostatic potential and density charge deformation (DCD) is reported as an effective method to account the charge transfer [46]. 2D MoSe$_2$ monolayer is a chemo resistive sensor that works on the change in resistance caused by the gain or loss of electrons between the gas molecule and 2D monolayer [48]. The charge transfer between the N$_2$O gas molecule and 2D MoSe$_2$ monolayer is bounded by the following equation as given in [48]:

$$Q_T = Q_{ads} - Q_{iso} \tag{2}$$

In this case, the charge transfer by the toxic N$_2$O gas molecule to the 2D MoSe$_2$ monolayer by varying the adsorbent distance is studied and shown in Table 2 and Fig.3 (a)-(b). It is observed from the Table 2 that as the target molecule moves closer towards the surface of 2D MoSe$_2$ monolayer, the charge transfer increases in both Top Mo and Top Se configurations subsequently. A value of $Q_T > 0$ indicates that N$_2$O donates electrons to the MoSe$_2$ monolayer whereas for $Q_T < 0$ MoSe$_2$ monolayer acts as donor to the N$_2$O gas molecule. In either case, there is change in resistance of the 2D MoSe$_2$ monolayer. For the Top Mo configuration, as the adsorbent height is decreased from 2 to 1.5 Å, an increase of 2.01% in charge transfer is noticed while from 1.5 Å to 1 Å it is 2.58 % which shows the robust chemisorption on the surface of 2D MoSe$_2$ monolayer. Likewise for the Top Se configuration, with a decrement from 2 to 1.5 Å the charge transfer increases by 2.79 % due to additional bond formation with the neighboring atoms, whereas from 1.5 Å to 1 Å the charge transfer exceeds significantly up to 9.42% as a result of strong binding. As the charge transfer is increased (decreased) the resistance of the 2D MoSe$_2$ monolayer decreases (increases) likely which is in justification with the working principle of the chemo resistive gas sensor.

**Table 2**

Values of Charge Transfer for different configurations at varying N$_2$O gas height.

| Configuration | Adsorbent distance (Å) | Charge Transfer (e) |
|---|---|---|
| Top Mo | 2 | 1.071 |
| Top Mo | 1.5 | 1.093 |
| Top Mo | 1 | 1.122 |
| Top Se | 2 | 1.112 |
| Top Se | 1.5 | 1.144 |
| Top Se | 1 | 1.263 |

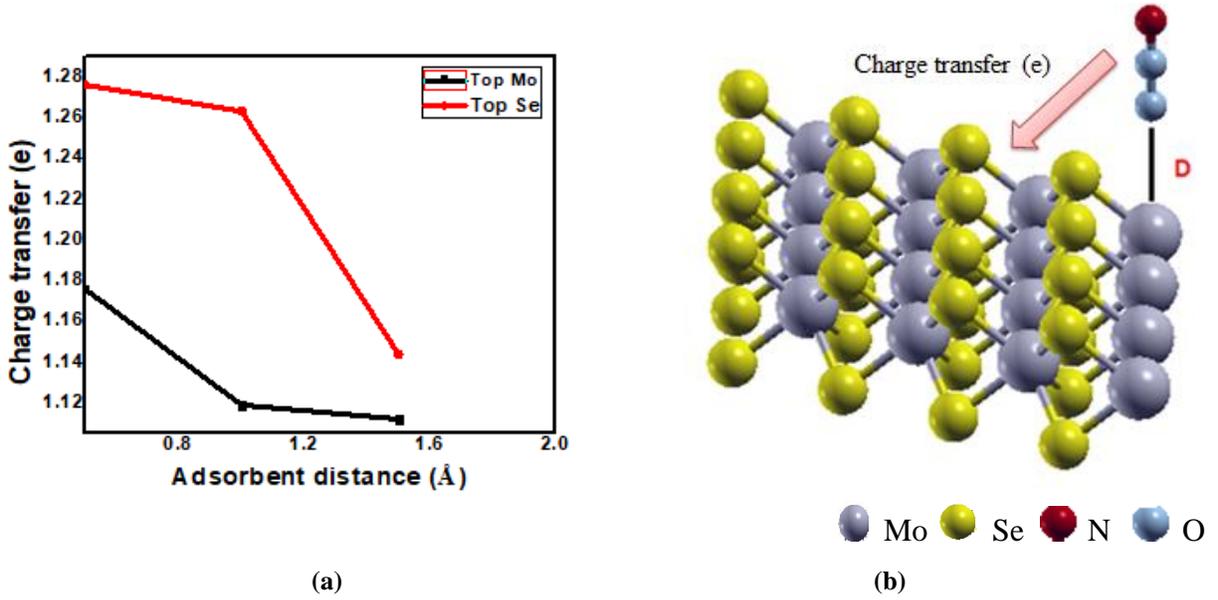

Fig. 3 (a) Charge Transfer variation with adsorbent distance for different configuration (b) $N_2O$ molecule as electron donor with varying adsorbent distance

Another response parameter which defines the behavior of the nano-gas sensor is its selectivity. Selectivity is defined as $\eta = (R_g - R_a)/R_a = \Delta R/R_a$ where $R_g$ and $R_a$ both are resistances under gas adsorption and in pristine state that is related to change in charge transfer in direct proportion as $\Delta Q/R_a$. Hence, the order of selectivity is found to be $\eta_{TopSe} \geq \eta_{TopMo}$ with order of resistance as $R_{TopMo} \geq R_{TopSe}$ respectively. Thus, for designing sensors with high selectivity it is required to have higher charge transfer between sensing material ($MoSe_2$) and analyte ($N_2O$) which is possessed by Top Se configuration.

*3.5 Recovery time: time span of toxic $N_2O$ gas molecule on 2D $MoSe_2$ monolayer*

The recovery time is another crucial parameter that estimates the surficial property of the 2D material. It defines the time of desorption of the exposed toxic $N_2O$ gas molecule from the surface of the 2D $MoSe_2$ monolayer [49]. In other words, it is the time span of gas molecule on the surface of 2D $MoSe_2$ monolayer [50]. The expression for the calculation of the recovery time is as follows [50]:

$$\tau = A^{-1} e^{\left(\frac{-E_a}{K_B * T}\right)} \quad (3)$$

Where $A^{-1}$ is the attempt frequency calculated to be $10^{-14}$ sec for $N_2O$ molecule (value varies depending upon the nature of gas molecule) [49], $E_a$ is the potential barrier equal to the value of $E_{ads}$, $K_B$ is the Boltzmann constant with a value of $8.62 \times 10^{-5}$ eV/K and T is the temperature (in K).

As obtained from above calculation of adsorption energy, it can be concluded that as lower the temperature, stiffer the gas desorption thus arising to a condition of higher adsorption energy. Interestingly, the time span of the gas molecule on the surface of the 2D $MoSe_2$ monolayer will be higher indicating to better chemisorption. Therefore, increase in temperature will be a better option that would quicken the process of desorption. Hence, lower recovery time for a nano-gas sensor favors it to be recyclable and contributing towards green technology [51]. The recovery time of 2D $MoSe_2$ monolayer for desorption of $N_2O$ gas molecule at different adsorption distances with varying temperature is presented in Table 3.

**Table 3**

Extracted recovery time of $N_2O$ gas molecule on 2D $MoSe_2$ monolayer at three diverse temperatures

| System | Adsorbent distance (Å) | Frequency Factor (A) | Temperature (K) | Recovery time (sec.) |
|---|---|---|---|---|
| Top Mo | 2 | $0.062 \times 10^{14}$ | 298 | $0.90 \times 10^{-14}$ |
| | | $0.082 \times 10^{14}$ | 398 | $0.237 \times 10^{-2}$ |
| | | $0.103 \times 10^{14}$ | 498 | $4.05 \times 10^{-3}$ |
| | 1.5 | $0.062 \times 10^{14}$ | 298 | $8.9 \times 10^{-32}$ |
| | | $0.082 \times 10^{14}$ | 398 | $3.44 \times 10^{-27}$ |
| | | $0.103 \times 10^{14}$ | 498 | $1.91 \times 10^{-24}$ |
| | 1 | $0.062 \times 10^{14}$ | 298 | $1.01 \times 10^{-39}$ |
| | | $0.082 \times 10^{14}$ | 398 | $3.87 \times 10^{-33}$ |
| | | $0.103 \times 10^{14}$ | 498 | $3.37 \times 10^{-29}$ |
| Top Se | 2 | $0.062 \times 10^{14}$ | 298 | $9.24 \times 10^{-14}$ |
| | | $0.082 \times 10^{14}$ | 398 | $1.06 \times 10^{-13}$ |
| | | $0.103 \times 10^{14}$ | 498 | $1.16 \times 10^{-13}$ |
| | 1.5 | $0.062 \times 10^{14}$ | 298 | $4.81 \times 10^{-14}$ |
| | | $0.082 \times 10^{14}$ | 398 | $6.51 \times 10^{-14}$ |
| | | $0.103 \times 10^{14}$ | 498 | $7.81 \times 10^{-14}$ |
| | 1 | $0.062 \times 10^{14}$ | 298 | $7.3 \times 10^{-15}$ |
| | | $0.082 \times 10^{14}$ | 398 | $1.59 \times 10^{-14}$ |
| | | $0.103 \times 10^{14}$ | 498 | $2.53 \times 10^{-14}$ |

It is clearly trending from Table 3 that with the increase in the adsorption distance of the target gas molecule towards monolayer, adsorption energy increases thus recovery time increases. To facilitate the process of desorption, temperature is increased from 298 to 498 K with a uniform step of 100 K, as a result of which recovery time is minimized from as shown in the Table 3. It is interesting to note that in both Top Mo and Top Se configurations, at adsorbent distance of 1 Å the adsorption energy is comparably high ranging from 1.55 (Top Mo) to 7.67 eV (Top Se) and does not supports feasibility in process of desorption thus increasing the recovery time $1.01 \times 10^{-39}$ to $3.37 \times 10^{-29}$ seconds for Top Mo and $9.24 \times 10^{-14}$ to $2.53 \times 10^{-14}$ considerably for Top Se configurations respectively. Thus, this clearly indicates that Top Se configuration is not supporting the desorption even at higher temperatures due to the higher adsorption energies resulting because of electron donating behavior of Se atoms whereas it is feasible for Top Mo configuration with comparatively lower adsorption energy shown due to electron accepting nature of Mo atom.

## 4. Conclusion

In this paper we extracted the structural and electronic properties of 2D MoSe$_2$ monolayer using first principle DFT theory in pristine and adsorbed state respectively. The adsorption of toxic N$_2$O gas on 2D MoSe$_2$ monolayer and its impact is analyzed through Top Mo and Top Se configurations respectively. The results show that the adsorption energies of N$_2$O gas molecule increase with the decrease in the adsorbent distance from 2 to 1 Å. Particularly, the highest adsorption energy of 7.67 eV indicates the extent of its interaction with the surface of 2D MoSe$_2$ monolayer which is shown by Top Se configuration. In addition, the charge transfer has significantly improved upon its expose to N$_2$O gas molecule. Upon the adsorption of the N$_2$O gas on MoSe$_2$, n-type semiconductor properties are determined as the Fermi level shifts upwards. Our calculations show that band gap of the 2D MoSe$_2$ monolayer is engineered from 1.45eV to 1.39(1.25) eV for both configurations in pristine and post adsorption respectively. Also, the magnetic moment is considerably influenced due to adsorption for both Top Mo and Top Se configuration respectively. At the same time, desorption performance of 2D MoSe$_2$ upon N$_2$O gas at three diverse temperatures is analyzed to prove it as a potential applicant. Top Mo configuration is shown to have smaller recovery time than the Top Se configuration turning it towards reusable nano-gas sensor. Our calculations shelter light on the application of 2D MoSe$_2$ monolayer as a potential sensor or disposer of toxic N$_2$O gas and open the area of designing environment friendly nano-gas sensors as a future scope of work.